\documentclass[prd,amsfonts,twocolumn,nofootinbib,showpacs]{revtex4}
\usepackage[english]{babel}
\usepackage[dvips]{graphicx}
\usepackage{enumerate}
\usepackage{color}

\def\sigmsd{\sigma_{SD}}

\def\sigv{\langle\sigma v\rangle}
\def\nuflub8{\phi^\nu_B}
\def\nuflube7{\phi^\nu_{Be}}
\def\epstrans{\epsilon_{trans}}
\def\epsann{\epsilon_{ann}}

\def\lesssim{\buildrel < \over {_{\sim}}}
\def\gtrsim{\buildrel > \over {_{\sim}}}

\begin{document}
\title{Effect of low mass dark matter particles on the Sun}
\author{Marco Taoso$^{1}$, Fabio Iocco$^{2}$, Georges Meynet$^{3}$, Gianfranco Bertone $^{2,4}$, Patrick Eggenberger $^{3}$}


\affiliation{$^{1}$ IFIC (CSIC-Universitat de Valencia), Ed.Instituts, Apt.22085, 46071 Valencia, Spain }
\affiliation{$^{2}$ Institut d'Astrophysique de Paris, UMR 7095-CNRS, Universit\'e Pierre et Marie Curie, 98 bis Boulevard Arago 75014, Paris, France} 
\affiliation{$^{3}$ Geneva Observatory, University of Geneva, Maillettes 51, 1290 Sauverny, Switzerland}
\affiliation{$^{4}$ Institute f\"ur Theoretische Physik, Universit\"at Z\"urich, Winterthurerstrasse 190, CH-8057 Z\"urich, Switzerland}

\begin{abstract}
We study the effect of dark matter (DM) particles in the Sun, focusing in particular on the possible reduction of the solar neutrinos flux due to the energy carried away by DM particles from the innermost regions of the Sun,  and to the consequent reduction of the temperature of the solar core.
We find that in the very low-mass range between 4 and 10 GeV, recently advocated to explain the findings of the DAMA and CoGent experiments, the effects on neutrino fluxes are detectable only for DM models with very small, or vanishing, self-annihilation cross section, such as the so-called asymmetric DM models, and we study the combination of DM masses and Spin Dependent cross sections which can be excluded with current solar neutrino data. Finally, we revisit the recent claim that DM models with large self-interacting cross sections can lead to a modification of the position of the convective zone, alleviating or solving the solar composition problem. We show that when the `geometric' upper limit on the capture rate is correctly taken into account, the effects of DM are reduced by orders of magnitude, and the position of the convective zone remains unchanged. 
\end{abstract}

\maketitle

\section{Introduction}

The identification of Dark Matter (DM) is one of
the most important open problems of modern cosmology and particle
physics. Among the many DM candidates proposed in the literature, 
Weakly Interacting Massive Particles (WIMPs) are the most popular, as they naturally achieve the appropriate relic density and
naturally arise in extensions of the Standard Model of particle physics that will soon be probed in accelerators.
WIMPs are currently searched for in a variety of experiments 
(see Ref.~\cite{reviews} for
recent reviews on particle DM, including a discussion of ongoing
direct, indirect and accelerator searches).

Here, we focus on the effect of WIMPs on stars, and more specifically on the Sun. 
In general, WIMPs crossing a star can in principle scatter off baryons to velocities
lower than the star escape velocity and be therefore gravitationally trapped
by the celestial body, the efficiency of the process
depending on the WIMPs scattering cross section and WIMPs ambient density.
Once captured, WIMPs can still scatter with the star nuclei and therefore transfer energy
inside the object. In addition, WIMPs can eventually annihilate, providing therefore
an exotic source of energy.
These effects have been investigated in the past, in particular
in the context of WIMPs 'cosmions' in order to solve
the solar neutrino problem \cite{cosmions,Dearborn:1990mm}.
This solution has then been discarded in
favour of neutrino oscillations and the combination of the cosmion
scattering cross-sections and masses suggested by the solar neutrino problem
is nowadays excluded by direct detection constraints.

Still, the intriguing possibility that WIMPs modify the properties
of the stars has recently prompted new interest on the subject.
Recent works have focused in particular on compact objects ~\cite{Moska, Bertone:2007ae},
main sequence stars at the Galactic center ~\cite{Fairbairn08,Scott:2008ns}
and first stars \cite{Ascasibar07,Freese,Freeseb,Iocco2,Freese2,Freese3,Iocco:2008rb,Taoso:2008kw,Yoon:2008km, RipamontiDM-star}.
In all these studies the celestial objects are placed in environments with high DM
density, thus enhancing the numbers of WIMPs captured by the star
and therefore the WIMPs transport and annihilation energy budgets.
However, in these scenarios it is not possible to set robust constraints on the DM cross sections and mass
because of the large uncertainties on the DM
densities and/or the lack of precise enough observations of the star targets.

The situation is quite different for the Sun since its properties have
been measured with good precision. Because of that, the Sun can be used
as a diagnostic tool to test small modifications of its structure 
induced by DM.
The most recent works in this direction have focused on modifications
of the solar neutrino fluxes and helioseismology data
\cite{Lopes:2001ig,Lopes:2002gp,Bottino:2002pd}.
Here we perform a complete and self-consistent calculation 
of the Sun evolution inside the galactic DM halo.
In particular we extend previous works considering light DM
candidates, with masses in the range suggested by DAMA \cite{Dama}
CoGent \cite{Aalseth:2010vx} and CDMSII \cite{Ahmed:2009zw} experiments.
We consider both standard annihilating WIMPs models and
scenarios with negligible DM annihilations.
We then move on Self Interacting Dark Matter (SIDM) models \cite{Spergel:1999mh},
which has recently been invoked as a solution of
the solar composition problem \cite{Frandsen:2010yj}.
Finally, we focus on Inelastic DM models \cite{TuckerSmith:2001hy}. 

The paper is organised as follows: in Section \ref{sec:theoretical} we introduce the DM models used in our calculations.
In Section \ref{sec:code} we present the GENEVA
stellar evolution code.
In Sec \ref{sec:diagnostictools} we discuss how the solar neutrino
fluxes can be used to constrain the DM parameter space.
We present our results for standard annihilating WIMPs in Section \ref{sec:pheno}, for
asymmetric DM models in Section \ref{sec:results}, for 
SIDM models in Section \ref{sec:self-inter} and for Inelastic DM in
Section \ref{sec:inelastic}.
Finally we summarize our conclusions in Section \ref{sec:concl}.
Technical details and the formulas implemented in the code
are in Appendix \ref{sec:formalism}.


\section{Theoretical setup and numerical code}
\label{sec:theoretical}

Galactic WIMPs inevitably stream through any celestial object; those which scatter
with a nucleus of the celestial object (in the following
we will limit ourselves to consider stars) loose energy in
such scatter, and if the energy loss brings the velocity
of the WIMP below the escape velocity, the WIMP is "captured", i.e. it becomes bound to the star.
The formalism to compute the rate at which WIMPs
are captured by a star has been object of extensive studies in the 
80's, and here we adopt the established results from 
\cite{Gould}.  The evolution of the total number of WIMP particles $N_{\chi}$ inside the Sun 
(or any celestial object) can be written as

\begin{equation}
\dot{N_{\chi}}=C-2A N_{\chi}^2- E N_{\chi} 
\label{diffequation}
\end{equation}
where C is the particle capture rate over the Sun, A is the annihilation rate and
E the evaporation one.
We discuss the details of the time evolution of the WIMP number density in the Appendix \ref{sec:formalism},
and we limit ourselves here to discuss the parameters entering in the calculation and the
implementation of the theoretical setup in the GENEVA numerical code for stellar evolution.

\subsection{WIMP parameters}

\paragraph{DM density.} The number of WIMPs captured by a star
depends on the WIMP scattering cross section off nuclei, 
ambient density and velocity distribution (see the Appendix for a detailed discussion).
The DM density distribution in the inner regions of the Galaxy is poorly constrained.
Semi-analytical studies and simple extrapolations of the
results of N-body simulations predict high DM densities at the center of galactic halos,
therefore suggesting stars at the galactic center \cite{Fairbairn08,Scott:2008ns}
and at the center or primordial halos
\cite{Freese,Freeseb,Iocco2,Freese2,Freese3,Iocco:2008rb,Taoso:2008kw,Yoon:2008km, RipamontiDM-star}
as promising targets to detect the effects of WIMPs on stars.

In this paper we focus on the Sun,
which being $\sim 8$ kpc distant from the galactic center, is
expected to be placed in a much lower DM density environment than the
galactic center.
Recent studies \cite{Catena:2009mf,Salucci:2010qr,Weber:2009pt}
have shown that current observations constrain
the local DM density, $\rho_{\chi},$ within a factor two, assuming
spherical DM profiles, while significant deviations may occur
for oblate DM distributions.
In the following, we adopt
$\rho_{\chi}=0.38$ GeV cm$^{-3},$ the value obtained
in \cite{Catena:2009mf} for the benchmark Navarro-Frenk-White density profile.
This value was obtained under the assumption that the DM profile is spherically symmetric, while recent numerical simulations that take into account the effect of baryons on the DM density profile suggest that the true DM density is at least $\sim 20$\% higher, and that it is affected by large systematic uncertainties due to the unknown position of the Sun in the triaxial DM potential of the Milky Way ~\cite{Pato:2010yq}.
Furthermore, the presence of a DM disc, corotating with stars and henceforth ``colder'' with
respect to the solar motion, could enhance the capture rate in the Sun up to
a factor ten \cite{Bruch:2009rp}. Our conclusions can be trivially rescaled for these cases,
which however we do not take into account for sake of conservativeness.

\paragraph{WIMP-nucleon scattering cross section.} The WIMP scattering cross section off nuclei is constrained by direct detection
experiments. The strongest bounds on Spin Indipendent (SI) interactions are currently set by
CDMS-II  \cite{Ahmed:2009zw}, XENON100 \cite{Aprile:2010um}
and CoGeNT \cite{Aalseth:2010vx} which impose
$\sigma_{SI}\leq 3 \mbox{ }10^{-40}-4 \mbox{ }10^{-42}$ cm$^2$ in the mass
range $m_{\chi}\sim 5-10$ GeV.
The strongest limit on Spin Dependent (SD) interactions is from PICASSO \cite{Archambault:2009sm}:
$\sigma_{SD}\leq 4 \mbox{ }10^{-36}- 4 \mbox{ }10^{-37}$ cm$^2$ in the same mass
range.
An evidence for an annual modulation signal has been reported
by the DAMA collaboration and subsequently confirmed by the upgraded
apparatus \cite{Dama,Damalibra,Bernabei:2010zza}.
A simple interpretation of this signal in terms of WIMP elastic
scattering off nuclei of the target material can hardly be reconciled with the constraints
from the other DM detection experiments. In this framework, in fact, the bulk
of the region of the $\sigma-m_{\chi}$ parameter space favored by
the DAMA signal is excluded by the null results of the other experiments,
both considering SI and SD interactions.

Still, there remain small surviving regions of the parameter space
for low WIMP masses, $m_{\chi}\sim 10$ GeV. We address the reader to
specific studies for more details (e.g. \cite{Savage:2009mk,Kopp:2009qt} and references
therein).
Recently, two experiments, CDMSII \cite{Ahmed:2009zw} and CoGeNT \cite{Aalseth:2010vx},
have reported an excess in their recoil energy spectra, even if the statistical significance
of these signals is very weak.
Interestingly, if interpreted in terms of elastic WIMPs scattering, these results
would point to light WIMPs, with masses and scattering cross-sections
close to the DAMA region \cite{Kopp:2009qt,Bottino:2009km}.
This scenario has been challenged by the very recent upper limits on SI interactions
obtained by XENON100 \cite{Aprile:2010um}.
However, there are subtle experimental effects that may cast some doubts on the low-mass end of this result, and as we write, the debate between the CoGent and Xenon collaborations is still open. 
Since the WIMP energy transport inside the star is indeed enhanced for light
WIMPs, in Sec.\ref{sec:results} we investigate the effects of these particles captured
by the Sun on its structure and observables.

\subsection{Dark matter models}

\paragraph{Standard WIMPs.}  Standard WIMPs achieve the
DM abundance inferred from cosmological observations
for $\langle \sigma v\rangle \sim3 \times 10^{-26} \mbox{ cm}^3\mbox{ s}^{-1}.$
Although significant deviations from this value are expected for example in presence of
Sommerfeld enhancements, efficient coannihilations or for
non standard cosmologies (see Ref.~\cite{reviews} for more details
and references),
deviations of few orders of magnitude
from the value quoted above are not relevant for the effects of
the WIMPs in the Sun, as we shall see in Sec. \ref{sec:formalism}. Indeed, the WIMPs annihilation and energy transport
are independent from $\langle \sigma v\rangle$ once an equilibrium
between scatterings and annihilations is reached. This occurs within a timescale
larger than the age of the Sun only for $\sigv \lesssim 10^{-33}$  cm$^3$s$^{-1}$,
after Equation \ref{eq:eqtime}.
Therefore, for the rest of the paper we fix
$\langle \sigma v\rangle = 3 \times10^{-26} \mbox{ cm}^3\mbox{ s}^{-1}$
when we consider annihilating WIMPs models.

\paragraph{Asymmetric dark matter}  

DM annihilations can be neglected either in models with very small or vanishing
annihilation cross sections or in presence of an asymmetry in the DM sector
between particles and anti-particles. 
This possibility, realized in the so-called asymmetric DM models,
is particularly interesting since these DM candidates can in principle 
have weak-scale interactions and therefore sizable scattering cross sections off baryons,
despite they do not annihilate.
Concrete realizations of this idea are for example models where the dark sector
contains a conserved $U(1)_X$ symmetry, analogue of the baryon number,
responsible for the stability of the lightest particle in the DM
sector. If this quantum charge is shared between baryons and DM it
can link the asymmetries in the two sectors and this may naturally explain why the baryons and
DM abundances are of the same order of magnitude
(see e.g.\cite{Hooper:2004dc,Kaplan:2009ag,Sannino:2008nv,Kribs:2009fy,Feng:2009mn}).
As noticed above, all DM models for which, the value of the annihilation cross section 
implies an equilibrium time $\tau_{\odot} > t_{\odot}$, namely 
$\sigv \lesssim 10^{-33}$  cm$^3$s$^{-1}$,
can be considered asymmetric from the point of view of WIMPs accretion.

\paragraph{Self-Interacting dark matter.}  
In the scenarios described before, the capture of WIMPs in the star is
induced by the interactions of DM with the nuclei.
In addition to that, also the elastic scattering of star crossing WIMPs
off previously captured WIMPs can contribute to the
DM capture rate.
For WIMPs self-interactions of the order of the WIMPs scattering
cross sections of baryons, like in the models previously considered, this contribution is
completely negligible since
the number of target nuclei in the Sun is much higher than
the number of trapped WIMPs.
However, this is not longer true for enhanced self interactions,
as predicted in Self Interacting Dark Matter (SIDM) models.
In Ref. \cite{Zentner:2009is} it has indeed been shown that for extreme values of self interaction
cross sections and for $\langle \sigma v \rangle \leq 10^{-27} \mbox{ cm}^3 \mbox{s}^{-1},$
the annihilation rate can indeed be boosted by a factor of order $\mathcal{O}(100)$, 
consequently enhancing the prospects for detections of 
high energy neutrinos produced by WIMPs annihilations
in the Sun.
SIDM models have been proposed in Ref. \cite{Spergel:1999mh}
to solve some discrepancies of the collisionless Cold Dark Matter (CDM) model, 
in particular the excess of substructures found in N-body simulations
of CDM halos with respect to observations and the conflict between the predicted cuspy CDM density profiles
and the observed cored halos of LSB galaxies.
Concrete particle physics realizations of SIDM models includes theories
with strongly interacting dark sector \cite{Sannino:2008nv,Kribs:2009fy,Feng:2009mn},
Q-balls \cite{Spergel:1999mh,Kusenko:1997si},
quark-gluino bound states \cite{Farrar:1984gk,Wandelt:2000ad}
and mirror DM models \cite{Mohapatra:2001sx,Berezhiani:2000gw,Foot:2004wz,Foot:2010th}.
Observations of galactic and galaxy clusters halos constraint
the size of DM self interactions. Here we report the
bound obtained from the bullet cluster as a robust upper limit
on the SIDM self interaction cross sections \cite{Randall:2007ph} : 
$\sigma_{\chi \chi}/m_{\chi} < 2 \times 10^{-24} \mbox{cm}^2\mbox{GeV}^{-1}.$
We address the reader to Ref. \cite{Taoso:2007qk}
for an updated discussion on the constraints on SIDM.

In the following, since the main focus of the paper is the effect of WIMPs on the
Sun, all stellar quantities we show are snapshots taken at the age of the Sun.

\section{The numerical code}
\label{sec:code}

In order to perform an accurate analysis of the problem we have implemented the capture, annihilation and energy transport of WIMPs in the GENEVA code
(for a description of this code, see \cite{Eggenberger:2008eg}). 

The implementation was done in the following way: from the structure of the stellar model, we can compute at each time step the number of WIMPs in the stellar model, the  energy released by annihilation and the energy transferred
by the WIMPs. At their turn, these energies modify the structure of the model. 
Thus the problem has to be resolved self-consistently.
To do that,
the rate of WIMPs annihilation and transferred energies are added in the equation describing the conservation of the radiative energy.
The code then, through an iterative procedure, looks for the values of the pressure, temperature, radius and luminosity  at different depths in the stellar model which match for the WIMP modified stellar structure equations (usually a stellar model consists of 900 layers from the surface to the centre).

With this numerical tool, we computed the evolution of one solar mass stellar models from the
Zero Age Main-Sequence (ZAMS) until the solar age (4.57 $\times$10$^{^9}$ years) 
for various prescriptions for the WIMPs energy transport and annihilation.
As initial composition we chose the following values: mass fraction of hydrogen, X=0.72, of helium, Y=0.266, and of the heavy elements, Z=0.014. The
distribution of the heavy elements was taken as given by \cite{Asplund:2005}. We did not include the effects of microscopic diffusion in these models, although this effect would
be required for the computation of tailored models for the Sun \cite[e.g.][]{Bahcall:1992ba}. The inclusion of this effect would however
not change the conclusions of the present work, which are deduced from detailed comparisons of models with exactly the same physics except for the inclusion or not of the WIMPs.
Since, microscopic diffusion would have to be accounted
in both models, it would not change in a significant way the relative effects obtained here.

\section{Diagnostic tools}
\label{sec:diagnostictools}

The modification of the stellar structure produced by the WIMPs induces changes in 
the frequencies of stellar oscillations modes and in the neutrino
fluxes. The first signature can in principle be observed with helioseismic
measurements \cite{Lopes:2001ig,Lopes:2002gp,Bottino:2002pd}, 
however, the neutrino flux is much more 
sensitive to the variation of temperature and density profile of the 
innermost regions of the Sun, and hence it is a much more powerful diagnostic
tool.
\begin{figure*}[t]
\includegraphics[width=0.45\textwidth]{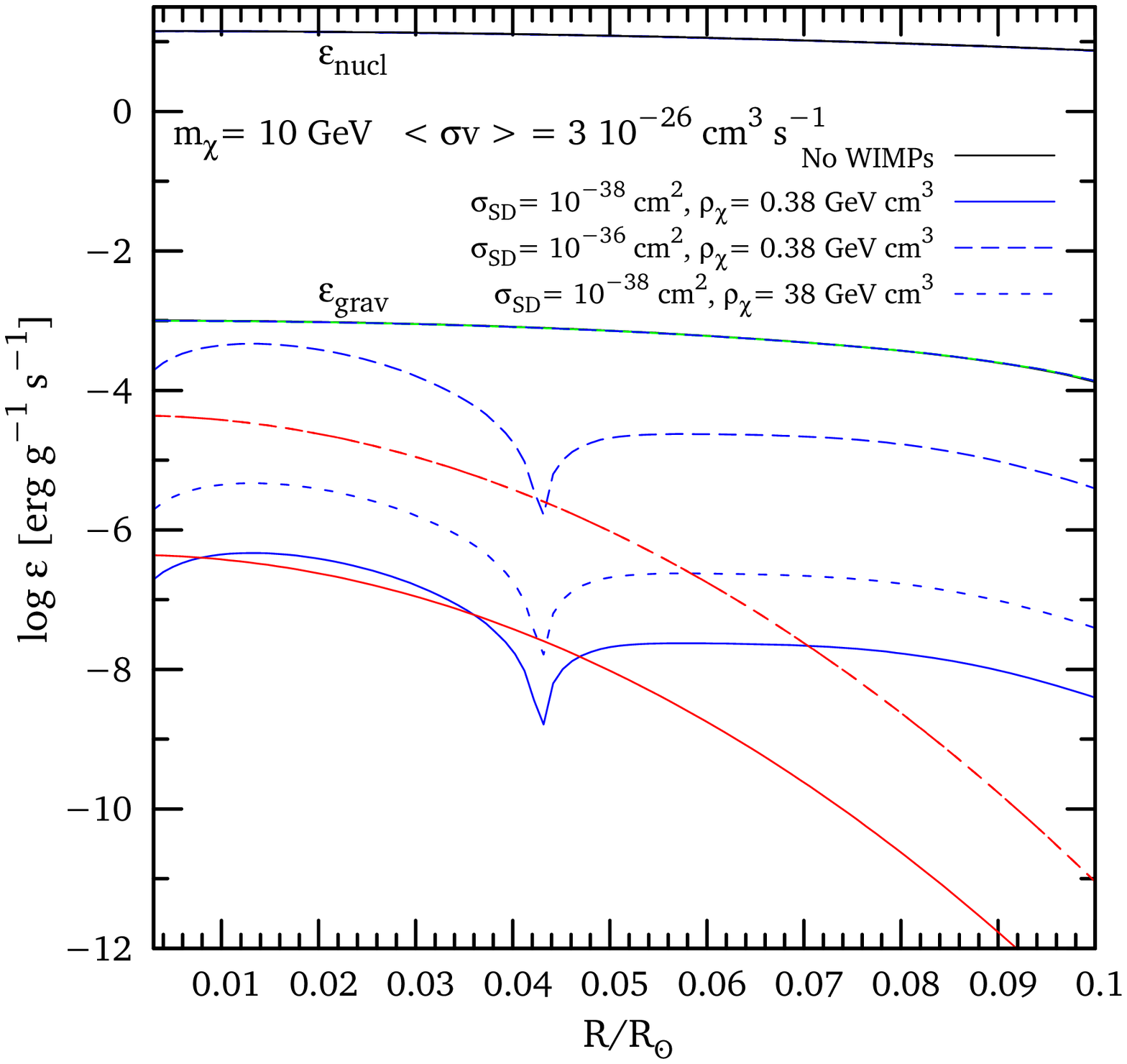}
\includegraphics[width=0.45\textwidth]{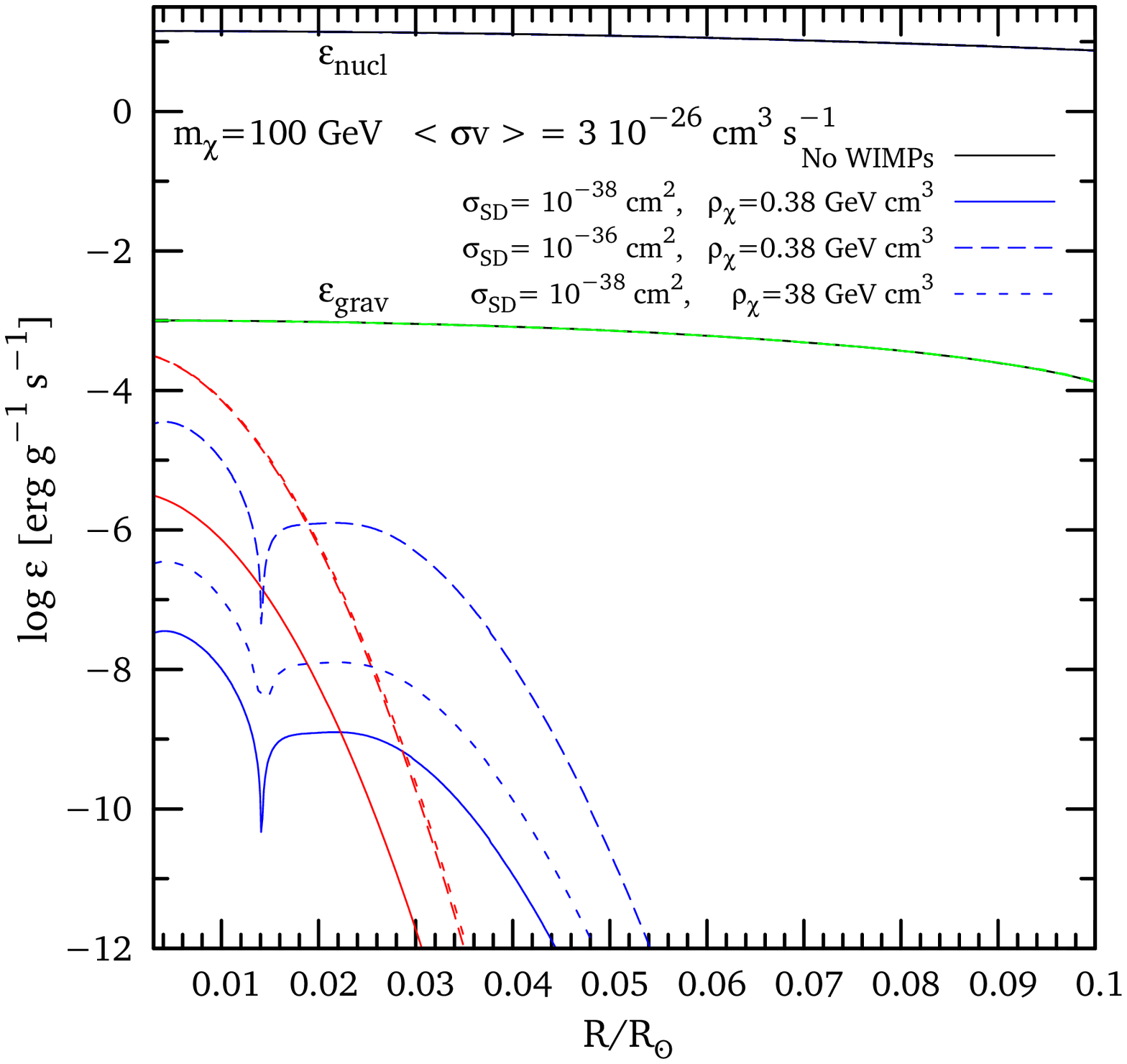}
  \caption{Comparison of the energy transported  ($\epsilon_{trans}$, blue lines) and injected (red) per unit mass and unit time by WIMPs in the Sun, with the nuclear $\epsilon_{nucl}$ and gravitational $\epsilon_{grav}$ energy density of the Star, for $m_{\chi}=10$ GeV (left panel) and $m_{\chi}=100$ GeV (right).
Here we plot the absolute value of
$\epsilon_{trans}$, the sign being negative at radii smaller than the dips
of the blue curves at $\sim0.04\mbox{ } R_{\odot}$ ($\sim0.014\mbox{ } R_{\odot}$) in the left (right) panel,
and positive otherwise. The sign of $\epsilon_{grav}$ is positive, this meaning a contraction of the star.}
  \label{fig:vanilaWIMPsunM10}
\end{figure*}

For the standard solar model, Ref. \cite{Bahcall:2002ba} indicates that the $^8${\rm B} neutrino flux varies as
$T^{25}$, therefore a 1\% temperature change will produce a 25\% change in the $^8${\rm B} neutrino flux.
However, as it has been in shown in Ref. \cite{Bottino:2002pd} , this simple scaling law is not valid to describe
the peculiar modifications of the temperature profile induced by the WIMPs.
Therefore, the use of a stellar code is mandatory to study the effects on DM on
the solar neutrino fluxes.

\begin{figure*}[t]
\includegraphics[width=0.32\textwidth]{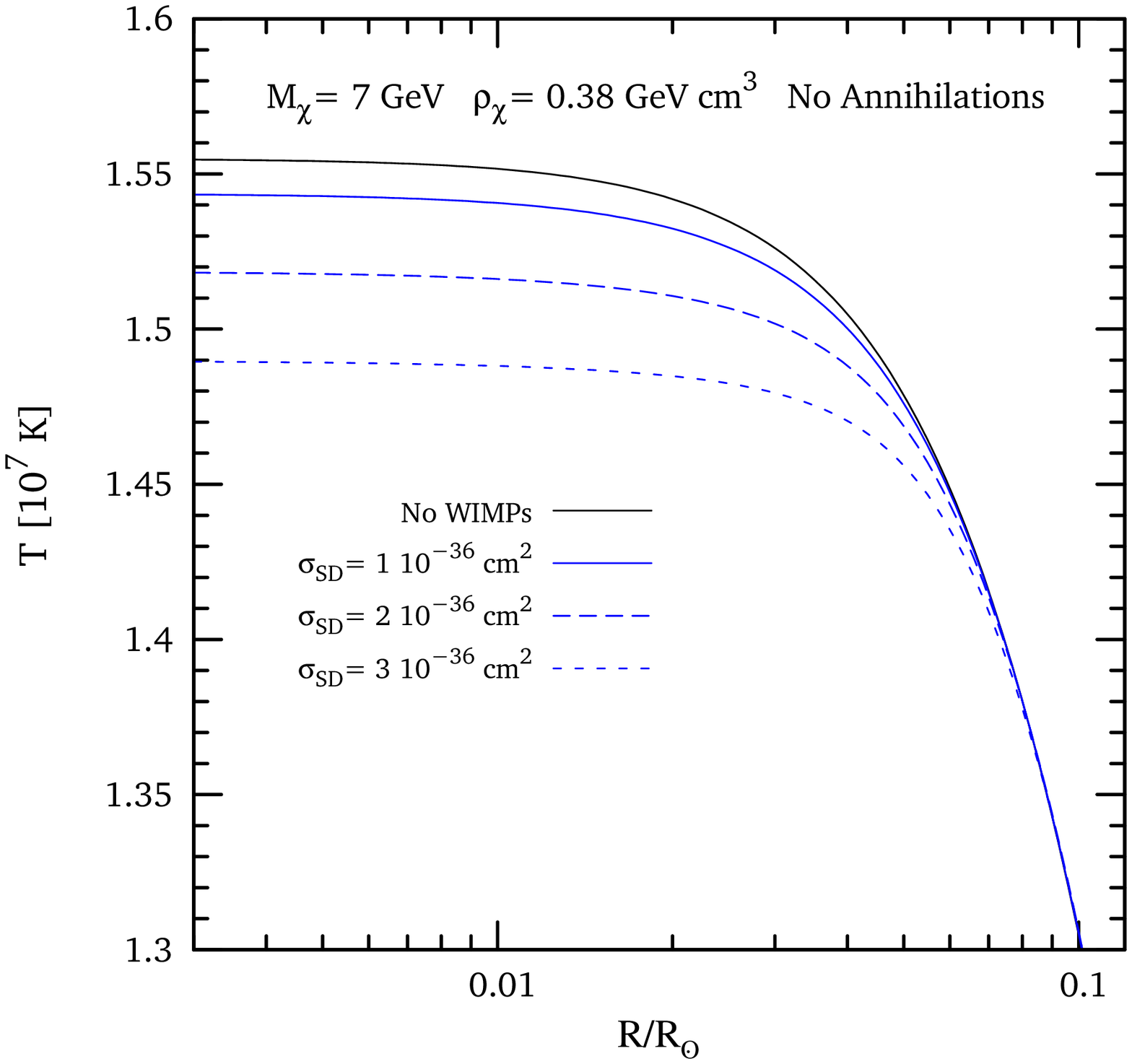}
\includegraphics[width=0.32\textwidth]{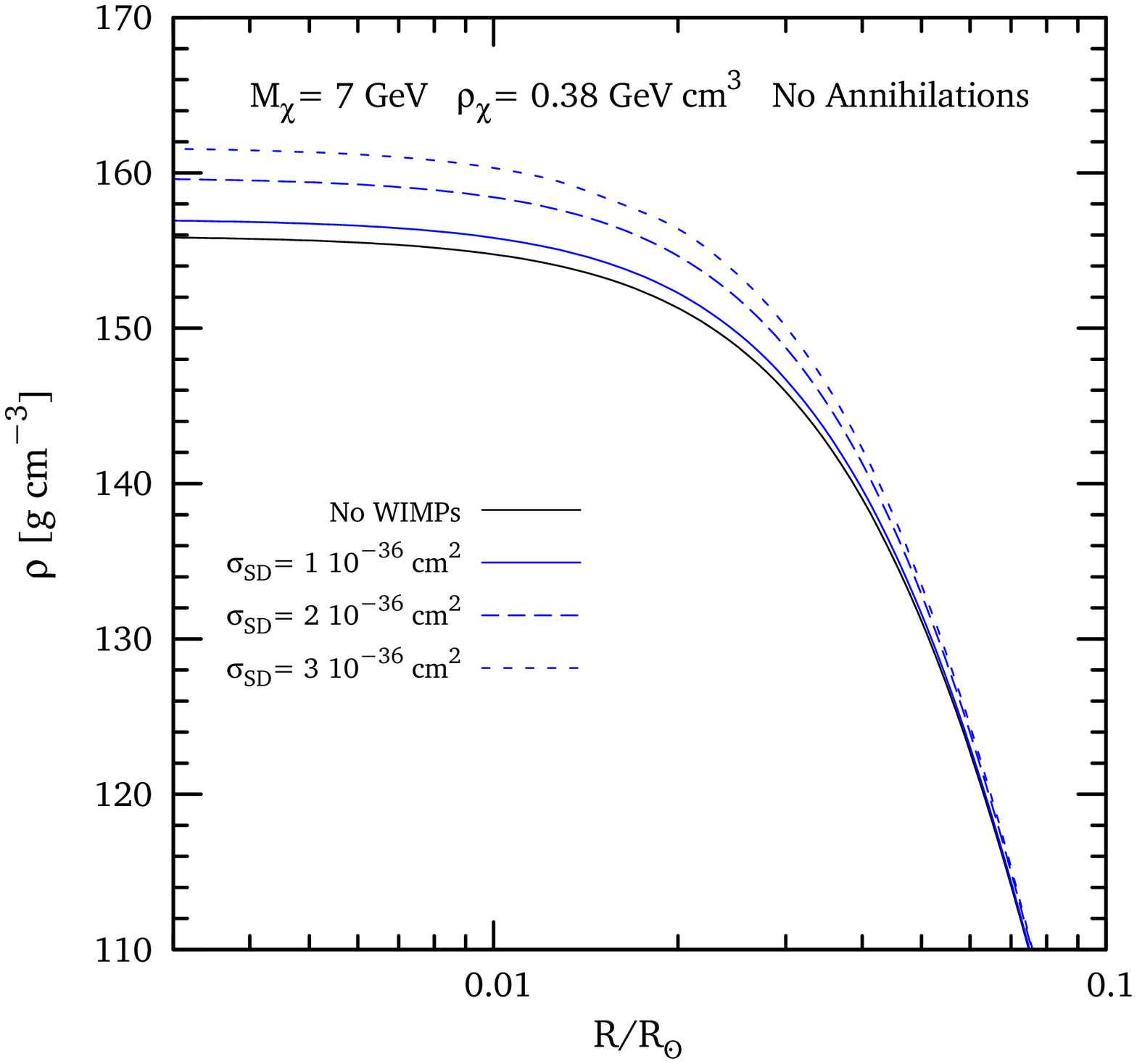}
 \includegraphics[width=0.32\textwidth]{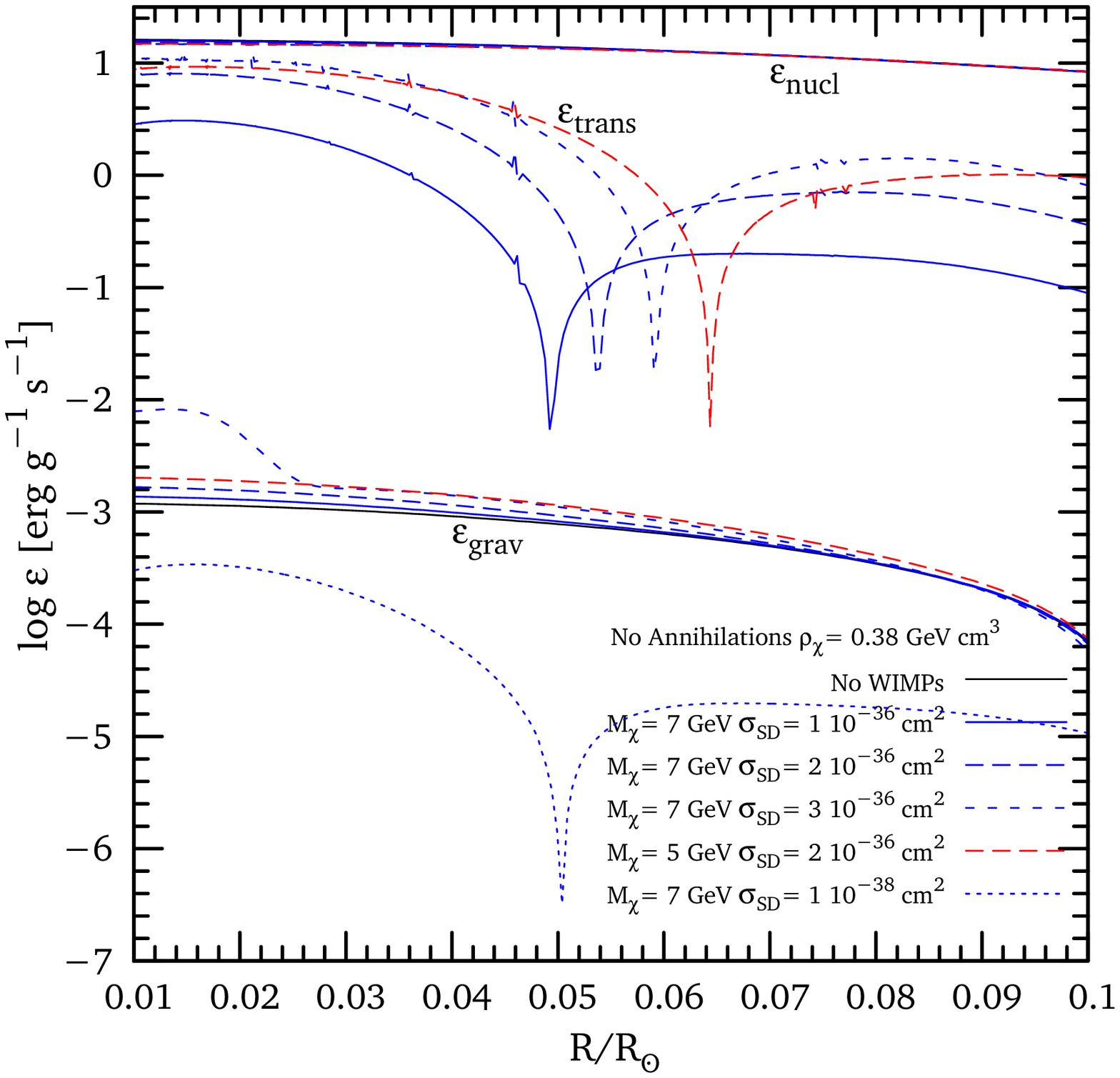}
  \caption{Impact of asymmetric DM on the Sun. {\it Left Panel: } Temperature profile of the Sun for different SD scattering cross-sections at $t_{\odot}=4.57$ Gyrs. {\it Central Panel: } Density profile (baryons only) inside the Sun. {\it Right Panel:} Energy transport (meaning of the curves as in Fig.\ref{fig:vanilaWIMPsunM10}). The sign of $\epsilon_{grav}$ is positive.}
  \label{fig:tempProf}
\end{figure*}

The distribution of WIMPs inside the Sun is crucial to determine the modifications
of the neutrino fluxes.
Once captured, WIMPs get redistributed inside a small spatial scale,
$r_\chi,$ of the order
of $10^9$ cm $\sim 0.01$ $R_{\odot},$ for a WIMP mass of $100$ GeV ($r_{\chi}$ scales
as $m_{\chi}^{-1/2}$, see Appendix \ref{sec:annihilation}).
The WIMPs energy transport will be therefore more efficient in the innermost regions of the
Sun core, as it can be appreciated in the left panel of Fig. \ref{fig:tempProf}
(see next sections for more details.)
The $^8${\rm B} and $^7${\rm Be} neutrinos, which are mostly produced
at $\sim 0.04 R_{\odot}$ and $\sim 0.06 R_{\odot}$, will be more affected by the presence of WIMPs
than the $pp$ neutrinos. In fact, although the $pp$ neutrinos are the most abundant solar neutrinos, they are mainly produced
at $\sim 0.1 R_{\odot},$ thus is a region well outside the one affected by the WIMP energy transfer.
For the same reason and considering the experimental uncertainties in the determination of 
$^8${\rm B} and $^7${\rm Be} neutrino fluxes, we conclude that
the $^8${\rm B} neutrino flux is the best diagnostic tool in order to test
the effects of WIMPs on the Sun.

The $^8${\rm B} flux has been determined with good accuracy
by SNO \cite{Aharmim:2009gd}:
$$\nuflub8=5.046^{+0.159}_{-0.152}\mbox{(stat)}^{+0.107}_{-0.123}\mbox{(syst) } 10^6 \mbox{ cm}^{-2}\mbox{s}^{-1}.$$
For completeness, we report also the $^7$Be neutrino flux inferred by Borexino\cite{Arpesella:2008mt}:
$$\phi_{Be}^{\nu}=(5.18 \pm 0.51)\times 10^9\mbox{ cm}^{-2}\mbox{s}^{-1}.$$
The solar model we use, as described in \ref{sec:code}, predicts at the solar age $t_{\odot}$:
$$\nuflub8=4.56 \times 10^{6} \mbox{ cm}^2\mbox{s}^{-1} \nuflube7=4.47 \times 10^{9} \mbox{ cm}^{-2}\mbox{s}^{-1} ,$$
values that are well in agreement with the experimental results to within the theoretical uncertainties of solar model calculations \cite{Serenelli:2009yc,PenaGaray:2008qe, Bahcall:2004mq}.

Despite its great success in explaining a large variety of observations,
the standard solar model suffers nowadays from the so-called {\it solar composition problems}.
Recent analysis points to a lower surface heavy element content than previously thought
(see \cite{Asplund:2009fu} for recent results)
and solar models incorporating these revised metallicities conflict
dramatically with helioseismological measurements, in particular right below
the solar convective envelope. With the solar abundances of \cite{Asplund:2009fu}, 
the radius of the convective zone, $R_{CZ},$ is more than 10 $\sigma$ higher than
the measured value: $R_{CZ}=0.713 \pm 0.001 R_{\odot}$
(see Ref.\cite{Serenelli:2009yc} for an analysis of comparison of different solar
models with helioseismology measurements.)
High and low heavy elements models also produce differences on $\nuflub8$ and  $\phi_{Be}^{\nu}$ 
respectively of $\sim 20$\% and $10$\%. 
However, the theoretical uncertainties
obtained within a particular solar model (with high or low metallicity estimations)
are significantly smaller, $\sim 10$\% and $6$\% for $\nuflub8$ and   $\phi_{Be}^{\nu}$ 
\cite{PenaGaray:2008qe, Bahcall:2004mq}.
Considering the theoretical and experimental uncertainties on $\nuflub8$,
we study the region of the DM parameter space where significant deviations
on $\nuflub8$ are found (the maximum allowed deviation of the $^8${\rm B} flux depends on the value
of the theoretical uncertainties considered. We define this thereshold in Sec.\ref{sec:results}).
As reference value, we consider the neutrino flux obtained from our solar model
without DM particles and in the Sec. \ref{sec:results} we define the maximum deviations
compatible with present data.

We now discuss the results for standard WIMPs, and provide a description of the role of WIMP energy transport and annihilation. We present separately in the next two sections the results for two special classes of models: asymmetric and self-interacting DM. 

\section{Results for Standard WIMPs}
\label{sec:pheno}

Since the trapped WIMPs are confined in small spatial scale inside the Sun, the annihilation of WIMPs in the star acts as
a point-like source of energy, whose effects
have been studied in details in the context of small, and high mass stars,
and for various initial metallicities (\cite{Fairbairn08,Scott:2008ns,Freese,Freeseb,Iocco2,Freese2,Freese3,Iocco:2008rb,Taoso:2008kw,Yoon:2008km,
Casanellas:2009dp}).
When the energy injected through DM annihilation is comparable with the nuclear one, a star finds its equilibrium
stage at lower temperatures, thus stopping in a colder region of the 
Hertzsprung-Russel diagram and prolonging its lifetime.
Although the effect is in principle interesting, the 
required DM densities and cross-sections are such that the effects
of an additional energy source due to the DM are totally negligible
on the Sun, and anywhere else in the Galaxy but in its innermost regions 
\cite{Fairbairn08,Scott:2008ns}.
The WIMPs in the star are confined into closed orbits and they inevitably
scatter off the stellar material therefore redistributing the energy inside the star.
The WIMPs mean free path is $l(r)=1/\sum_i  \sigma_i n_i(r)$,
where the sum is performed over all the target nuclei,
$\sigma_i$ is the WIMP scattering cross section off a certain nucleus and
$n_i(r)$ is the nucleus number density profile in the star.
For reference, the WIMP mean free path is typically
of the order $\sim$10$^{10}$cm for $\sigma_{SD}\sim 10^{-36} \mbox{ cm}^2.$
Once compared with the typical radius of the WIMP cloud
$r_\chi$=10$^9$ cm, this 
indicates that WIMP transport effects are non local.
In our computations we span many orders of magnitude on the
WIMP scattering cross sections so we investigate both local
and non-local regimes.
We address the reader to Ref. \cite{Scott:2008ns}, and 
references therein for further details. The equations implemented in the GENEVA code
are summarized in the Appendix \ref{sec:formalism} for sake of completeness.

In the non local regime regime, WIMPs scatters can efficiently transport 
the heat from the inside of the stellar out
to colder regions, 
thus operating toward a flattening of the temperature profile, and a 
consequent readjustement of the entire stellar structure.
Raising the WIMP-proton scattering cross section $\sigma_0$, the capture rate (modulo the caveat on saturation
of the optical limit, see the Appendix \ref{sec:capture}) and the WIMPs scattering rate grows,
therefore increasing the energy transported by the WIMPs, $\epstrans$.
The dip in the behavior of $\epstrans$ in Figure~\ref{fig:vanilaWIMPsunM10}
corresponds to the change in sign of $\epstrans$:
at small radii WIMPs are ``colder'' than the baryons, and acquire energy with their scatters,
which they deposit outward.
Since the mean free path of WIMPs scales like $l \sim 1/\sigma_0,$
at large scattering cross sections the WIMPs remain progressively ``trapped'' in the interior
of the star.
This implies that the heat transport from DM particles becomes local and therefore
$\epstrans$ dramatically reduces.

The WIMP mass also plays a role in determining the effects on
DM in stars, as it can be appreciated in Equations in Appendix \ref{sec:formalism}.
Lowering the DM mass goes in the direction of maximizing the transport effects, but also
the evaporation rate. Above
m$_\chi$= 5 GeV, evaporation can be safely neglected
and the mass of the particle acts to modify the radius and normalization
of DM inside the star.

In Figure~\ref{fig:vanilaWIMPsunM10} we compare
the rate of energy released per unit of baryonic mass by DM annihilation $\epsann$ and the rate of energy absorbed or released by
DM transport $\epstrans$, with
the rate of energy produced by nuclear reactions.
The solid lines represent two benchmark WIMP scenarios for the Sun: 
a local DM density of
$\rho_{\chi}$=0.38 GeV/cm$^3$, a Spin-Dependent
scattering cross section $\sigmsd$=10$^{-38}$cm$^2$
and WIMP mass and self-annihilation
cross-section compatible with a DM thermal production scenario.
Since in these computations the WIMP annihilations and transport energies 
are negligible with respect to
the energy provided by nuclear reactions, the structure of the Sun is not affected
by the presence of the WIMPs. 
As a consequence of that, the solar neutrino fluxes are unchanged with respect to their values
in our standard solar model without DM particles.
As stated before, the WIMP energy transport is more efficient for light WIMPs:
in Figure~\ref{fig:vanilaWIMPsunM10} it dominates over the annihilation energy for a 10 GeV WIMP
while the opposite trend occurs for a 100 GeV WIMP,
at least in the inner region of the Sun.
However, due to the different scaling of $\epsilon_{ann}$ and $\epsilon_{trans}$
with $\rho_{\chi}$, at high DM densities the annihilation energy is always the most relevant
source of energy produced by the WIMPs in the star.
An example of that is given by the dashed curve in Figure~\ref{fig:vanilaWIMPsunM10},
where we raise the DM density of two orders of magnitude, 
$\rho_{\chi}$=38 GeV/cm$^3$.

Rescaling our curves in Fig.~\ref{fig:vanilaWIMPsunM10} using Eqs. \ref{diffequation},
 \ref{eq:captrate} and \ref{eq:epsann},
it can be noticed that energy injection by DM annihilation is comparable with the nuclear energy
for DM densities of order $\rho_{\chi}\sim$10$^8$GeV/cm$^3.$ 
This value is indeed the typical DM density
for which the behaviour of main sequence stars starts to be severely modified by 
WIMPs annihilations \cite{Fairbairn08, Scott:2008ns}.

In conclusion, the Sun's global properties are not
dramatically changed by self-annihilating WIMPs, and no
solar observable can be used as a good diagnostic for these models.
As we explain in the following Sections, things are quite
different for other DM scenarios.

\section{Results for Asymmetric DM}
\label{sec:results}

If WIMPs are asymmetric, Equation \ref{diffequation}
reads: 

\begin{equation}
\dot{N_\chi}=C-E N_{\chi},
\end{equation}
and particle will continue accumulating in the center of the star, since their abundance will not be limited by annihilation.

As before, DM particles rapidly thermalize, therefore the shape of the WIMPs distribution stays as in the annihilating case whereas the normalization
$N_{\chi}$ will result modified.
The number of WIMPs inside the Sun, neglecting evaporation, will simply
be $N_{\chi}=C t_{\odot},$ with $t_{\odot}$ the age of the Sun. In the annihilating case,
after a transient $\tau_{\chi}$, an equilibrium between annihilations and capture is
reached and $N_{\chi}$ stays $N_{\chi}=C \tau_{\chi}.$
Comparing $t_{\odot}=4.57\times 10^9\mbox{ yr}$ with the typical equilibrum timescale
in a vanilla WIMPs scenario, $\tau_{\chi}\sim 10^6 \mbox{ yr } (\sigv/10^{-26}\mbox{ cm}^3\mbox{s}^{-1})^{-1/2}$
for $m_{\chi}=10$ GeV, it is evident that the number of WIMPs trapped in
the Sun is significantly bigger in asymmetric models.

In order to study the effects of asymmetric DM, we have computed the variations of neutrino fluxes produced by the heat transport
of WIMPs in the Sun. We evolved the Sun from the ZAMS up to its current age $t_{\odot}$
with the GENEVA stellar code. As previously discussed, the WIMPs effects are
accounted for self-consistently at each time-step; if, for instance, energy is evacuated by the WIMPs
in the very central layers, the structure will react by allowing the central layers to contract. 
The present numerical approach
is therefore more accurate and self consistent than methods in which the effects of WIMPs are
deduced from the structure of standard solar models computed without WIMPs effects,
this, for at least two
reasons: first because in the present approach the structure of the star can readjust itself to any changes
produced by the WIMPs and second because, these readjustments are accounted for during the whole previous
nuclear evolution of the Sun.

The first two panels of Figure \ref{fig:tempProf} shows the temperature
and density profiles of the Sun in presence of asymmetric DM. 
Note the change in temperature with increasing SD 
scattering cross section: the temperature decreases at the center and
(although this is difficult to appreciate in the plot) slightly increases
close to the external edge of the stellar core.
The reason of the decrease in temperature can be understood in terms of energy transported away from the solar core.
In general, any amount of energy removed from the core leads to its contraction.
Naively, one would expect a warming of the internal regions of the Sun since 
part of the energy extracted from the gravitational energy reservoir 
goes into internal energy.
However, the energy carried away by the WIMPs is well above the energy released by the core contraction, 
resulting therefore in a cooling of the central regions.
 
The right panel in Figure  \ref{fig:tempProf} demonstrates in fact that the energy transported by asymmetric DM can become extremely high, 
and it can significantly modify the structure of the star. We show the results for two WIMP masses: at fixed $\sigmsd$ the normalization of $\epstrans$
is higher for a 5GeV particle than for a 7GeV one, and the sign inversion
point of the energy transport term shifts outwards. 
The shift in the sign inversion point between different curves for the same DM mass is due to the feedback induced by the modification
of the density profile, which starts becoming sensible at $\sigmsd\approx$10$^{-36}$.

\begin{figure*}[t]
\includegraphics[width=0.45\textwidth]{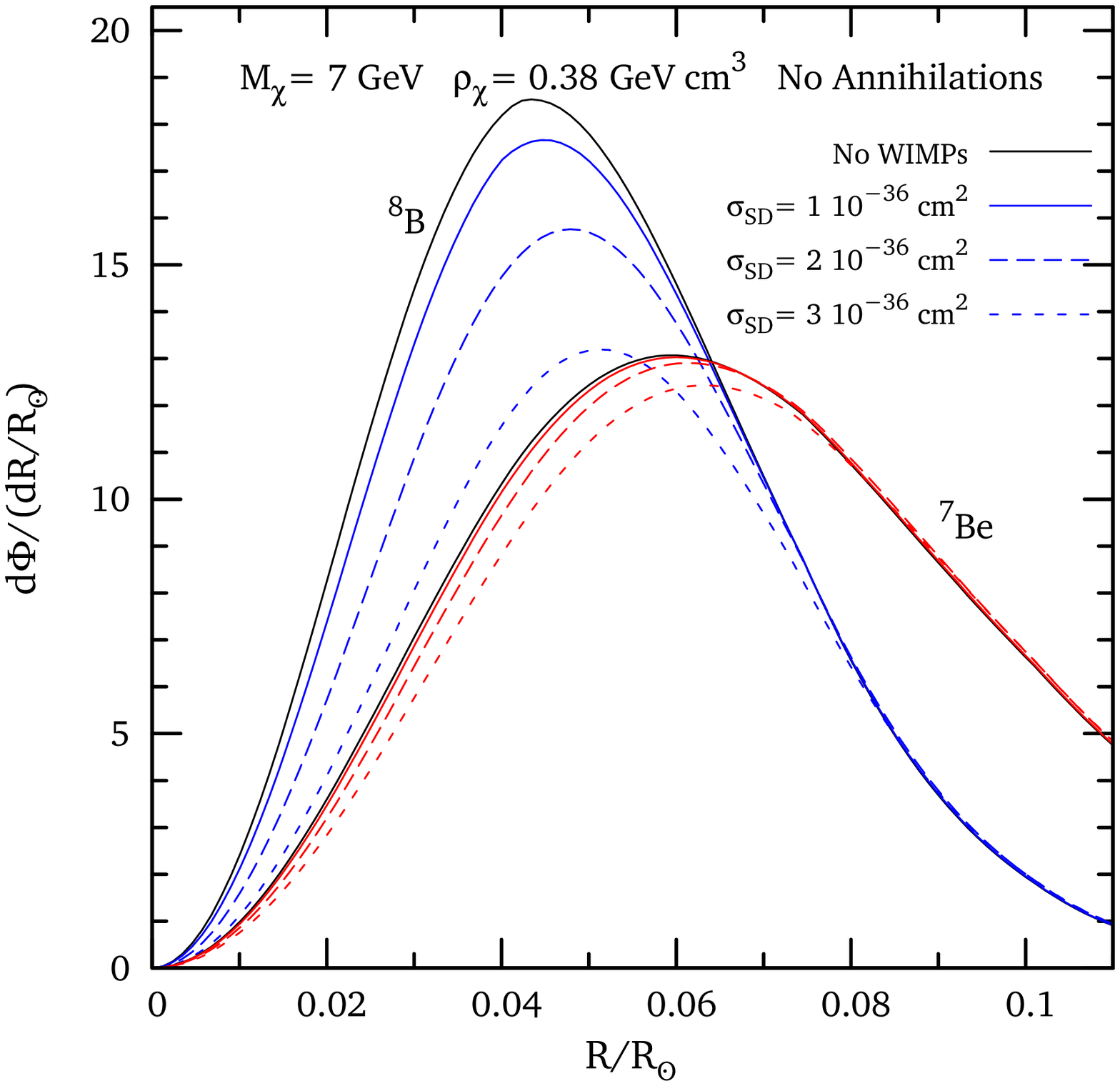}
     \includegraphics[angle=0, width=0.45\textwidth]{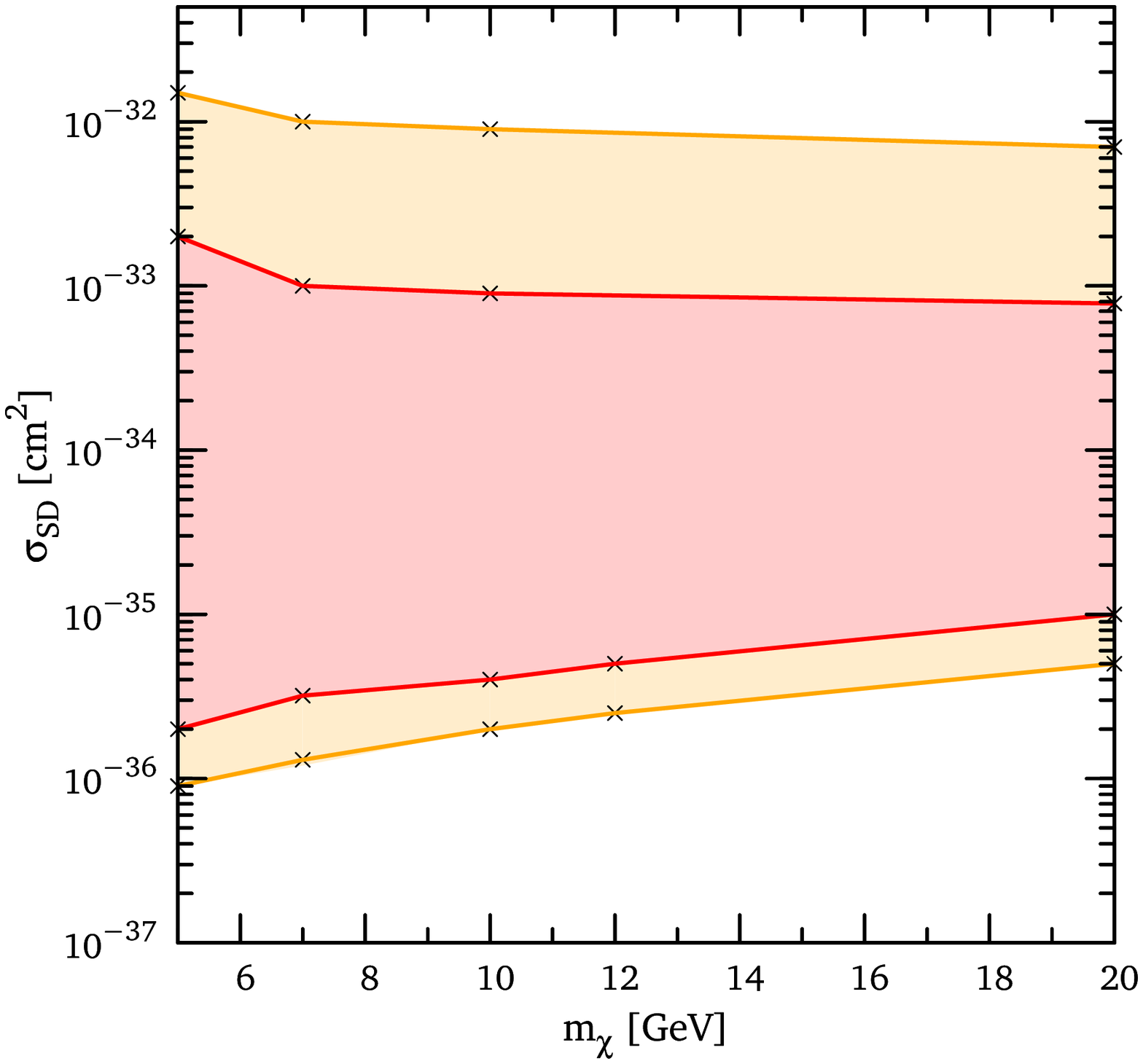}
  \caption{Solar neutrino fluxes in presence of asymmetric DM in the Sun. Left Panel: Differential $^8$B and $^7$Be neutrino fluxes as a function of radius. The curves referring to our solar model are normalized to unity. 
Right panel: isocontours of 25\% (red) and 5\% (yellow) $\nuflub8$ deviations with respect our solar model
prediction in the $m_{\chi}$-$\sigma_{SD}$ plane. The red area show the region of the parameter
space with $\nuflub8$ modifications larger than 25\% and therefore in tension
with present $\nuflub8$ data.
The weakening of the DM transport effects at high cross sections is due to the transition to local transport regime. See text for details.
}
  \label{b8evol}
\end{figure*}

In principle one should study how the structural modifications 
modify the details of neutrino oscillations in the Sun and consequently the
inferred values of the neutrino oscillations parameters, 
i.e. $\Delta m^2_{\odot}$ and $\sin(\theta_{\odot}).$
A sizeable effect is actually expected for density variations of the order of $10$\% 
at $\sim 0.04 R_{\odot},$ which corresponds to the region of maximal $^8$B production \cite{Bolanos:2008km}.
However, at that radii, we find a decrease of the density below $2$\% even for models 
producing modifications of the $\nuflub8$ larger than 20-25\%.
This modification is comparable with the uncertainties on the density profile
in the inner regions of the Sun, which is of the order of $1\%$ for
$R<0.45 R_{\odot}$ \cite{Bahcall:2005va}.

These considerations are valid also for the $^7$Be fluxes and therefore modifications of the neutrino
oscillation probabilities are completely negligible for our purposes.

Left panel of Figure~\ref{b8evol} shows the differential $^8${\rm B} and $^7{\rm Be}$ neutrino
fluxes as a function of the stellar radius,
in presence of different WIMP models.
As expected, the reduction of the neutrino production, due to the cooling
of the baryons inside the Sun, is more efficient
at small radii, where the WIMPs are concentrated.
As noticed in Sec. \ref{sec:diagnostictools}, 
this leads to a larger modification on the total $^8${\rm B} neutrino flux,
$\nuflub8$, which is the integral over the whole Sun of the corresponding differential 
quantity plotted in the left panel of Figure~\ref{b8evol}, than those on the $^7{\rm Be}$
neutrino flux, $\phi_{Be}^{\nu}$.

To study the impact of these structural variations on the solar neutrino flux, 
we have performed a systematic study of the DM parameter space, varying the WIMPs scattering cross section
and mass. For SI interactions we find sensible variations of the neutrino fluxes
only for very large values of $\sigma_{SI}$, already severely excluded
by direct detection experiments, therefore we focus for the rest of the section on SD interactions.

In the right panel of Fig. \ref{b8evol}, we show in the $m_{\chi}-\sigma_{SD}$ plane the isocontours 
corresponding to $\nuflub8$ variations of 25\%
and 5\% with respect to our solar model without WIMPs.
For masses below $4-5$ GeV the evaporation becomes important and the number of WIMPs
inside the star is strongly suppressed.

Above $m_{\chi}=20$ GeV the
WIMP transport starts to become inefficient and even for high scattering cross
section (still in the non-local transport regime) 
the WIMP energy transport is non negligible only at the very center of the star,
providing a local dip of energy.
However, for increasing WIMPs masses the existing constraints 
from direct detection experiments become
more severe, and the region of the parameter space able to produce
sizeable modifications of $\nuflub8$ is already excluded. Because of that, we do not
explore that region any further.
At very high scattering cross sections (i.e. $\gtrsim$10$^{-33}$cm$^2$) 
the WIMPs heat transport becomes more and more
localised and the modifications of $\nuflub8$ tend to decrease, consistently with what
is described in Sec.\ref{sec:pheno}. This is also the reason of the non
specularity of the exclusion curves in Figure \ref{b8evol}: the low and high cross-section
regions are characterized by different physics (non-local vs local transport
effects, respectively).

Combining experimental and conservative (20\%)
theoretical uncertainties we derive that modifications of the $\nuflub8$
above $\sim 30$\% are excluded at $95$\% CL.
The maximum modifications that we obtained in our computations are slightly below
this level: further increasing
$\sigma_{SD}$, problems are encountered in solving the stellar structure at ages well below the solar one. 
We have not further investigated if these difficulties are
merely a numerical artifact or are instead related to the non existence
of a solution of the stellar structure.
However, we notice that once we obtain variations of $\nuflub8$ of the order of
$20$\%, further small changes of $\sigma_{SD}$ induce rapid modifications
of $\nuflub8$. Because of that, the
 isocontours corresponding to $30$\% $\nuflub8$ variations
should be closed to the ones corresponding to $\delta \nuflub8=25$\%
shown in Fig. \ref{b8evol}.

Considering a more optimistic value for the theoretical uncertainties
on the $\nuflub8$ predictions, i.e. 10 \%, the threshold for exclusion at 95\% CL is lowered
to 18\% variations from our solar model prediction.
Most of the SD cross-sections inside the $25$\% region in Fig.\ref{b8evol} are
excluded by the direct detection experiments constraints, apart from a small region at low masses
which is somewhat in tension with those bounds.
A reduction of the theoretical and experimental uncertainties on
$\nuflub8$ in the next years may in principle improve the sensitivities
on $\sigma_{SD}.$
We show however in Fig.\ref{b8evol} that 
considering a 5 \% modification of $\nuflub8$ the region
of the $m_{\chi}-\sigma_{SD}$ parameter space which can be probed enlarges
very little.
Our findings are qualitatively consistent with previous results \cite{Dearborn:1990mm,Bottino:2002pd}.

\section{Results for Self-interacting DM}
\label{sec:self-inter}
As we have seen, since DM annihilations limit the number of 
trapped WIMPs in the Sun, asymmetric DM models
are promising scenarios to look for modifications of the Sun properties.
This is particularly true for the case of SIDM models, as it has been noticed
in Ref. \cite{Frandsen:2010yj}.
In presence of only DM self-interactions and DM scattering
off baryons, the number of DM particle inside the Sun is given by:

\begin{equation}
\dot{N_{\chi}}=C +C_{\chi\chi} N_{\chi} 
\label{diffequationSIDM}
\end{equation}

where $C_{\chi\chi}$ is the DM capture rate through DM self interactions.
The solution of this equations reads:
\begin{equation}
N_{\chi}(t)=\frac{C}{C_{\chi\chi}} \left( e^{C_{\chi\chi} t}-1 \right)
\label{numberSIDM}
\end{equation}
and reduces to $N_{\chi}(t)=C t$ for negligible
self interactions, i.e. the formula used in Sec.\ref{sec:results} 
It can be noticed therefore that the numbers of
captured DM particles is exponentially enchanced in asymmetric SIDM models.

The self-capture rate $C_{\chi\chi}$ has been obtained in
Ref \cite{Zentner:2009is}:
\begin{equation}
C_{\chi \chi}=  \sqrt{\frac{3}{2}} \rho_{\chi} \frac{\sigma_{\chi \chi}}{m_{\chi}} \frac{v^2(r)}{\bar{v}}
\langle \phi_{\chi}\rangle \frac{\mbox{Erf}(\eta)}{\eta}.
\label{captureSIDM}
\end{equation}
where the average of $\phi(r)\equiv v^2(r)/v^2(R_*)$ over the WIMPs distribution gives
$\langle \phi_{\chi}\rangle\simeq 5.1.$
$R_*$ refers to the Sun radius and the definition of the other quantities
can be found in Appendix.

An upper bound to the total WIMPs self-interaction
arises when the sum of the short range WIMPs self-interaction cross sections over all
the WIMPs targets equals the surface of the WIMPs distribution:
$$\sigma_{eff, \chi}\equiv \sigma_{\chi \chi} N_{\chi} = \pi r_{\chi}^2$$
In the equation above, we can safely consider that all the WIMPs are localized
inside the thermal readius $r_{\chi}.$
Once the geometrical limit is saturated at a time $\hat{t}$
with a number of WIMPs $N(\hat{t})$ inside the star,
we simply replace the combination $\sigma_{\chi \chi} N_{\chi}$
in Eq.\ref{diffequationSIDM} with $\sigma_{eff, \chi}$.
The WIMPs number density for $t>\hat{t}$ is therefore given by
$$\hat{N}_{\chi}(t)=\left( C+\hat{C}_{\chi \chi} \right)\left(t-\hat{t}\right)+N(\hat{t})$$
with $\hat{C}_{\chi \chi}$ obtained replacing 
$\sigma_{\chi \chi}$ in Eq. \ref{captureSIDM} with $\sigma_{eff, \chi}$.

\begin{figure}[t]
  \resizebox{\hsize}{!}{\includegraphics[width=0.45\textwidth]{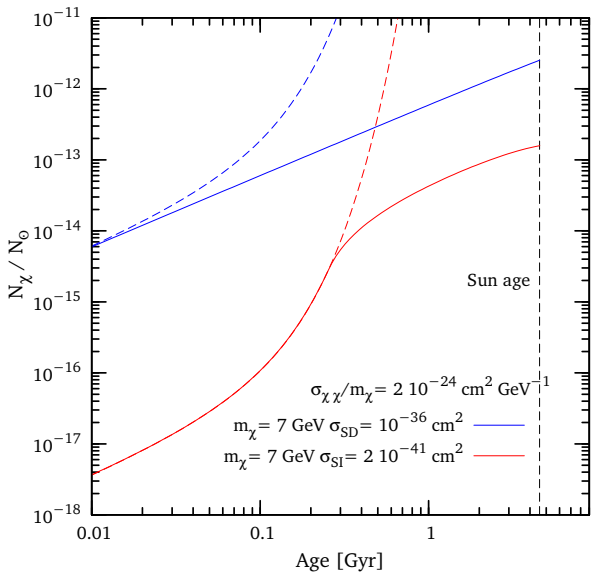}}
  \caption{Number of WIMPs inside the star normalized to the number of baryons as
a function of the age in the asymmetric SIDM scenario. Solid and dashed lines refer to calculations
which respectively implement or neglect the  geometrical bound on the self interaction.}
  \label{nwimps}
\end{figure}

We have studied the effect of a population of SIDM models in the Sun,
implementing in the GENEVA code equations \ref{diffequationSIDM}--\ref{captureSIDM}.
To exemplify our results we focus on a DM mass $m_{\chi}=7$ GeV
and we consider $\sigma_{SI}=2 \times 10^{-41} \mbox{ cm}^2$
and $\sigma_{SD}=10^{-36} \mbox{ cm}^2$, i.e. values close to the bounds
set by direct detection constraints.
In Fig.~\ref{nwimps} we show the evolution in time of the number of captured WIMPs normalized to
the total number of baryons, $N_{\chi}/N_{\odot}$ for the two models here considered and
for a self-interaction cross sections which saturates the bound from
the bullet cluster mentioned above.
In the case of a pure SI interaction it can be noticed the change
between the exponential rise regime and the linear one, occurring when
the geometrical limit is saturated.
For SD interactions the same happens but since the geometrical bound
is obtained earlier in the evolution this modification is difficult to notice.
In both calculations we have not found any significant deviation of
the solar structure at the Sun age.
For the case of SD interaction, the WIMPs transport energy $\epsilon_{trans}$
is at most a factor 2 smaller than the nuclear energy.
Small radiative opacity variations occur only at $r\lesssim0.05 R_{\odot}$.
These small effects does not lead to any significant variations of
the sound speed profiles in the whole Sun , and so of $R_{CZ}.$

In Ref.\cite{Frandsen:2010yj},
adopting a polytropic model for the Sun structure and the linearized
solar model of Ref. \cite{Villante:2009xs} the authors concluded that for suitable values of 
SIDM parameters the boundary of the solar
convective zone $R_{CZ}$ is decreased.
They claimed that these modifications are such that the solar composition problem mentioned above
can be solved, i.e. the position of the convective zone and 
helioseismology data can be reproduced.

Our analysis does not confirm their conclusions.
In order to maximize the WIMP energy transport, we have artificially turned off
the geometrical bound in our code and evolved different solar models.
We stress however that this scenario is unphysical.
We have studied both models with pure SD and pure SI interactions,
leaving the DM mass and cross section at the values fixed before.
We have then varied the DM self-interaction cross-section in order to maximize
the number of SIDM in the Sun.
In the extreme cases we have found
a dramatical reduction of the core temperature, which produces a
significant decrease of $\nuflub8$, in agreement with what is obtained
in Sec.~\ref{sec:results}.
At the center of the star, the luminosity is reduced, due to the evacuation of energy
produced by the WIMPs, and the radiative opacity is increased, as a result of the 
decrease of the temperature and the increase of the density.
The external zones are not significantly affected by the WIMPs
so we do not see any significant change in the position of the convective zone.
We conclude that a population of WIMPs, being strongly localized at the center of the Sun,
can not affect its external shells without at the same time completely change
the internal structure of the star.
On the other hand, strong modifications of the central solar structure
traduce in dramatic changes on the solar neutrino fluxes and
helioseismology g-modes, as also demonstrated in Ref. \cite{Cumberbatch:2010hh},
and in general of any observable sensitive to the physical conditions in that inner
regions.
The results of Ref. \cite{Cumberbatch:2010hh} appear in good agreement with our own and confirm
that the presence of WIMPs inside the Sun can not modify the solar sound
speed profile in such a way to restore the agreement with helioseismological
data.

\section{Inelastic DM}
\label{sec:inelastic}

The possibility that DM particles could scatter inelastically
up to an excited state, has been recenly proposed
in the literature in the framework of theoretically justified 
models \cite{TuckerSmith:2001hy,Thomas:2007bu,ArkaniHamed:2008qn,Cui:2009xq}, in order
to explain the modulation observed in the DAMA signal. 
The existence of two states splitted by an energy $\delta$
introduces a threshold for the WIMPs scattering since only WIMPs with
enough kinetic energy, $E_k$ can be excited to the heavier state:
\begin{equation}
E_k \geq \delta \left( 1+\frac{m_{\chi}}{M_{N}}\right)
\label{eq:inelastic}
\end{equation}
with $m_N$ the mass of the nuclei.
Since the energy threshold depends on the type of nucleus considered,
in the inelastic DM scenario the scattering rate can differ dramatically between
different targets, opening the possibility to reconcile the DAMA signal
with the results of the other experiments \cite{Kopp:2009qt,Chang:2008gd,MarchRussell:2008dy,Arina:2009um,SchmidtHoberg:2009gn}.
Recent analyses favor WIMP masses of the order $m_{\chi}\sim10-50 \mbox{ GeV}$
and small mass splitting $\delta \sim 30-130 \mbox{ KeV}.$
We address the reader to the original literature for more details.

The WIMPs capture rate in the Sun for inelastic DM models
tends to be reduced with respect to the elastic case by the presence of the inelastic barrier.
However, since a significant fraction of the kinetic energy 
lost by the WIMPs goes into excitation, less energy is transferred
to the nuclei, thus reducing the nucleus form-factor suppression.
Explicit calculations have shown than, for certain choices of the DM parameters,
the capture rate can be enhanced up to an order of magnitude \cite{Nussinov:2009ft,Menon:2009qj}.
Another important effect of the existence of the energy threshold
$\delta$ for scatter is the modification of the internal distribution
of DM particles inside the star: as a consequence of the reduced
number of scatters, they can't shed away their angular momentum
entirely, and their orbits remain larger than in a standard, elastic
DM case (see Fig 1a in \cite{Nussinov:2009ft}).

Here we aim to understand if inelastic DM models can 
affect the Sun's properties, and if the diagnostics we have discussed in
previous Sections allows to probe this class of models.
First, we notice that, once captured, the velocity of DM particles will be smaller than the local escape velocity.
Considering an escape velocity at the Sun core of $v=1300 \mbox{km s}^{-1}$ and taking a
WIMPs speed at infinity $u=220 \mbox{km s}^{-1}$
the maximum kinetic energy of a trapped WIMPs will be $E_k^{max}\sim 96.5 \mbox{ KeV} (m_{\chi}/100 \mbox{ GeV})$.
Looking at Eq. \ref{eq:inelastic}, it can be realised that 
once captured, WIMPs can scatter only with elements heavier than hydrogen, even
considering a small mass splitting $\delta =30$ keV.
Simple kinematic considerations show that after a few scatterings the WIMPs will
not have enough kinetic energy to overcome the inelastic threshold. For example,
for the same $\delta=30 \mbox{ keV},$ a WIMP with mass $m_{\chi}=100 \mbox{GeV}$ will scatter in average $\sim$ 5 times
off iron or off helium.
Therefore, at a given time, only a small
part of the WIMPs population inside the star can effectively scatter
and transport energy.
Focusing on the most optimistic asymmetric case,
the number of WIMPs above the kinematic
thereshold, $N^{\prime}_{\chi},$ at given time can be obtained from:

\begin{equation}
\dot{N^{\prime}_{\chi}}=C-N^{\prime}_{\chi} \tau_{in}^{-1} 
\label{eq:inelasticdiff}
\end{equation}

where $\tau_{in}$ is the average time needed for a WIMP to fall below
the inelastic thereshold.
This quantity can be estimate as:

\begin{equation}
\tau_{in}= N_{sc} \frac{l}{w}
\label{eq:inelastictau}
\end{equation}
where $l$ is the WIMP mean free path defined in Sec. \ref{sec:pheno},
$w$ is the WIMP velocity inside the star and $N_{sc}$ is the number of WIMP scatterings.
Here we take a mean baryon density $\rho \sim 10^2 \mbox{ g cm}^{-3}$ and
we consider only scattering off elements heavier than helium, which in total
account for $\sim 1\%$ of the Sun mass.
We obtain therefore:
\begin{equation}
\tau_{in}= 2.3 \times 10^{5} \mbox{ s} \frac{N_{sc}}{30} \frac{1318 \mbox{km s}^{-1}}{w} \frac{100 \mbox{g cm}^{-3}}{\rho}
\frac{10^{-36} \mbox{ cm}^2}{\sigma_{SD}}
\label{eq:inelasticnum}
\end{equation}

At the age of the Sun, the number of WIMPs transporting energy inside the star will
be $C \tau_{in},$ which is many orders of magnitude lower than in the elastic case,
$C t_{\odot}.$

Considering the capture rate obtained with the most favored values of inelastic
DM parameters, from the results obtained in Sec. \ref{sec:results}
we conclude that inelastic DM models do not produce any detectable
modifications of the Sun properties.

\section{Conclusions}
\label{sec:concl}

By the use of a stellar evolution code, we have studied the modifications on the Sun structure induced by 
DM particles captured by the Sun during its lifetime.
For standard WIMPs an equilibrium between the capture and
the annihilation rates is reached in short timescales.
After this transient period, the number of WIMPs inside the Sun is constant and we find it to be too small
to produce any observable effect on the Sun properties.

However, for scenarios with null or very small annihilation cross sections,
$\sigv \lesssim 10^{-33}$  cm$^3$s$^{-1},$ the number of trapped WIMPs is significantly increased.
In this case, the transport of energy by the WIMPs from the interior of the Sun
to the outer shells can dramatically reduce the core temperature,
the most important consequence being a reduction of the $^8B$ flux.
Considering the present theoretical and experimental uncertainties
on $\nuflub8$, we have studied the combination of
DM masses and SD scattering cross sections which can be ruled our with this argument,
finding only a small region of the parameter space which is not already excluded by
direct detection bounds.
Even with a significant decrease of the uncertainties on $\nuflub8$, the region
of the parameter space which can be probed remains approximately the same, therefore future experimental advances will not significantly change the situation.

We have also considered SIDM candidates which have been recently invoked as
a solution of the solar composition problem.
Correctly implementing the geometrical limit, we find that the number of
WIMPs captured on the star is not sufficient to produce significant deviations
on the Sun properties, unless one considers very high scattering cross sections, where however 
the models become in tension with direct detection bounds.
Even in this case, we encounter significant modifications of the density
and temperature profiles only in the inner regions of the Sun while
the outer shells are not significantly affected.
We can therefore exclude that the transport of energy produced by a population
of WIMPs can solve or even alleviate the solar composition problem.

Finally, we notice that the WIMPs energy transport may produce dramatic effects on the structure of stars settled in high DM density environments, like the galactic center.
We leave the investigation of this scenario for future work. 

When this manuscript was ready for submission, a similar work appeared on the arXiv,
focusing on the effects of WIMPs in the Sun \cite{Cumberbatch:2010hh}.
While our work mainly focus on the modifications of the solar neutrino fluxes,
Ref. \cite{Cumberbatch:2010hh} considers future helioseismology data as a diagnostic tool to constrain the WIMPs parameter space.

\acknowledgments 
We thank N.Fornengo and A.Palazzo for useful conversations.
We thank S.Sarkar and M.T.Frandsen for discussions and the comparison
of the results.
MT acknowledges support
of the Spanish MICINNs Consolider–Ingenio 2010 Programme under
grant MULTIDARK CSD2009-00064. His work
is partly supported by the Spanish grants FPA2008-00319 (MICINN) and PROM-
ETEO/2009/091 (Generalitat Valenciana) and the European Council (Contract Number
UNILHC PITN-GA-2009-237920). FI acknowledges support 
from the European Community research program FP7/2007/2013 
within the framework of convention \#235878.

\appendix

\section{Formalism and code}
\label{sec:formalism}

\subsection{Capture}
\label{sec:capture}
The formalism to compute the rate at which WIMP particles
are captured by a star have been extensively studied in the 
eighties. Here we adopt the results from 
\cite{Gould}, who gives for the capture rate $C$:

\begin{equation}
C= \sum_i 4 \pi \int_{0}^{R_*} dr r^2 \frac{dC_i(r)}{dV} \label{eqn:C}
\end{equation}
with
\begin{eqnarray}
\frac{dC_i(r)}{dV} &=&  \left(\frac{6}{\pi}\right)^{1/2}
\sigma_{\chi,N_i} \frac{\rho_{i}(r)}{M_i}\frac{\rho_{\chi}}{m_{\chi}}
\frac{v^{2}(r)}{\bar{v}^2} \frac{\bar{v}}{2 \eta A^2} \\
\nonumber &\times & \left\{ \left( A_+ A_- -\frac{1}{2}\right)
[\chi(-\eta,\eta)-\chi(A_-,A_+) ] \right.\\ \nonumber &+& \left.
\frac{1}{2} A_+ e^{-A_-^2} -\frac{1}{2} A_- e^{-A_+^2} -\frac{1}{2}
\eta e^{-\eta^2} \right\} \label{eqn:dCdV}
\label{eq:captrate}
\end{eqnarray}
$$ A^2=\frac{3 v^2(r)\mu}{2 \bar{v}^2 \mu_-^2} \mbox{,  }\hspace{0.5cm}
A_{\pm}=A \pm \eta \mbox{,}
\hspace{0.5cm}\eta^2=\frac{3v_{*}^2}{2\bar{v}^2}$$
$$\chi(a,b)=\frac{\sqrt{\pi}}{2}[\mbox{Erf}(b)-\mbox{Erf}(a)]=\int_a^bdy e^{-y^2}$$
$$\mu_-=(\mu_i-1)/2 \mbox{,} \hspace{0.5cm} \mu_{i}=m_{\chi}/M_i$$
where $\rho_i(r)$ is the density
profile of a given chemical element in the interior of the star and
$M_i$ refers to its atomic mass, while $m_{\chi}$  and $\rho_{\chi}$
are respectively the WIMP mass and the WIMP density at the star position. 
The analytic expression for the capture rate per shell volume reported above, $dC/dV,$
is obtained for the case of a Maxwell-Boltzmann velocity distribution
with speed dispersion $\bar{v}$;
throughout this paper, we adopt $\bar{v}=270  \mbox{ km s}^{-1}$,
and we take the velocity of the star moving through the DM halo,
 labeled as $v_*$, to be $v_*=220 \mbox{km s}^{-1}$, as appropriate for the Sun. 
 The radial
escape velocity profile depends on $M(r)$, i.e. the mass enclosed within a
radius $r$, $v^2(r)=2 \int_{r}^{\infty} G M(r^{'})/r^{'2} dr'$.
The DM elastic scattering cross section off nuclei $\sigma_{\chi N_i}$
is the sum of the Spin-Dependent (SD) and Spin-Independent (SI) contributions:
$$\sigma_{\chi N_i}=\beta^2 A_i^2 \sigma_{SI}+\beta^2 \sigma_{SD} \frac{4 (J_i+1)}{3 J_i} 
 |\langle S_{i,p} \rangle + \langle S_{i,n} \rangle |^2 $$
with $\sigma_{SI}$ and $\sigma_{SD}$ the hydrogen-normalized
SI and SD nuclear-scattering cross sections.
The factor $A_i$ is the atomic number while $J_i$ is its spin
and $\beta$ the ratio of the reduced mass of the WIMP-nucleus
and WIMP-proton systems.
Finally, $\langle S_{i,p} \rangle $ and $\langle S_{i,n} \rangle $
are respectively the expectation values of the spin content of the
proton and neutron group in the nucleus.
The total capture rate is then obtained summing up all the elements
present in the star.

The geometrical size of the star itself sets an upper bound to the effective
WIMP scattering cross-section,$\sigma_{\mbox{eff}},$ which we model 
imposing a maximum to the capture rate once the condition below is reached:
$$\sigma_{eff}\equiv\sum_i \sigma_{\chi N_i} N_i = \pi R_{\star}^2$$
with $N_i$ the number of nuclei of the element $i$; this is sometime
referred to as the optical limit.

\subsection{Annihilation}
\label{sec:annihilation}

Throughout this paper we work under the assumption that WIMPs
thermalize inside the Sun within a negligigible timespan with respect to
stellar evolutionary timescale. This is a good approximation for most WIMP models,
and especially for the high scattering cross-sections we will examine,
as an upper limit of the thermalization time reads \cite{Iocco:2008rb}:
\begin{equation}
\tau_{th}=\frac{4 \pi}{2 \sqrt{2G}}\frac{m_{\chi}}{\sigma_{\chi N_i}}\frac{R_*^{7/2}}{M_*^{3/2}}
\label{eq:thtime}
\end{equation}
that for the sun means
$$\tau_{th}\sim 10^5 s\mbox{ } \left(\frac{m_{\chi}}{10 \mbox{ GeV}} \frac{10^{-36} \mbox{ cm}^2}{\sigma_{\chi N_i}}\right).$$

In the case of a non-local WIMP energy transport, i.e. for large $K$ (see next section),
the WIMPs clould shares a global temperature which can be set, in good approximation,
to the star core temperature $T_c$.
The WIMP radial distribution in the star gravitational well
is thus given by \cite{GriestSeckel}:
\begin{equation}
n_{\chi, nl}(r)=n_{\chi,0} e^{\frac{-r^2}{r_{\chi}^2}}, 
r_{\chi}=\sqrt{\frac{3 k T_c}{2\pi G \rho_c m_{\chi}}}
\label{eq:Distribuzione}
\end{equation}

\noindent with $\rho_c$ referring to the core density.
The distribution results quite concentrated toward the
center of the star: typical values for ZAMS stars are 
$r_\chi\sim$0.1 $r_c\sim 10^{9}$ cm, $r_c$ being the stellar core radius.
Such zero order distribution is slightly modified by transport effects,
see Section \ref{sec:transport}, and we adopt the modified distribution
to compute the actual rate of annihilations; however, the effects of such
modification are small and their effects on annihilations negligible in practice.
Once the normalization $n_{\chi,0}$ is obtained solving the Eq.\ref{diffequation},
the distribution $n_{\chi}(r)$ is completely specified and the
annihilation term can be easily computed:
\begin{equation}
A= \int_0^{R_\odot} \epsann r^2  4 \pi \rho(r) dr ,
\label{eq:annihil}
\end{equation}
with:
\begin{equation}
\epsann=\frac{1}{2}\sigv m_\chi c^2 n^2_{\chi}(r) \rho(r)^{-1}.
\label{eq:epsann}
\end{equation}

$\rho(r)$ is the baryon density at a given
position and $\epsann$ is the luminosity produced by WIMPs annihilations
per unit of baryonic mass.
The factor 1/2 (1/4) in the equation above is 
appropriate for self (not self) conjugate particles
and $\sigv$ the velocity-averaged annihilation cross-section.
If an equilibrium between capture and annihilation is reached
the annihilation rates reduces to A=C/2 and it is independent
on the annihilation cross-section.
The equilibrium timescale for such a processes, 
neglecting evaporation, is given by
\begin{equation}
\tau_{\chi}=\left(\frac{1}{C A}\right)^{1/2}\sim\left(\frac{\pi^{3/2}r_{\chi}^3}{C \sigv}\right)^{1/2}
\label{eq:eqtime}
\end{equation}

and for annihilations cross-sections of the 
order $\langle \sigma v\rangle \sim 10^{-26} \mbox{ cm}^3\mbox{ s}^{-1},$ 
a typical value suggested by the requirement of the correct
WIMP relic density, it is much shorter than the
age of the Sun.
In our calculations we compute the capture and
annihilation terms self-consistently, and
find good agreement with the equilibrium approximation for this value
of the annihilations cross-section.
However, we also explore scenarios with negligible or null annihilation
cross-sections, as for the case of asymmetric-WIMPs, for which the equilibrium
between annihilations and capture is not reached.

\subsection{Evaporation}
\label{sec:evaporation}

WIMPs trapped inside the Sun may scatter off nuclei to velocities
high enough to escape the gravitational field of the star and thus
leave the system.
Detailed studies on this phenomena \cite{Gould:1987ju,Gould1987b}, called evaporation, 
have shown that
the  evaporation timescale, defined as the inverse of
the evaporation rate, $E$ in Eq.\ref{diffequation}, depends exponentially
on the mass of the WIMPs.
Therefore, below a certain mass threshold,
$m_{ev},$ basically all the WIMPs evaporate during the Sun evolution
and conversely, slightly above $m_{ev}$ the evaporation is negligible.
At high WIMP scattering cross sections, corresponding to the LTE regime,
the evaporation starts to be inefficient since the WIMPs leaving the center of the star
are rapidly re-scattered to low velocity orbits.
The evaporation mass, depends therefore of the size and nature (SI or SD)
of the scattering cross section.
Considering the Sun, for $\sigma_{SD}\sim 10^{-36} \mbox{ cm}^2$ it is typically of the order
$m_{ev}\sim 4 \mbox{ GeV}$ and it decreases to $m_{ev}\sim 2 \mbox{ GeV}$ 
at $\sigma_{SD}\sim 10^{-34} \mbox{ cm}^2.$
In the following we focus to WIMP masses $m_{\chi} \geq 5 \mbox{ GeV}$
so the effect of the evaporation can be completely neglected for our purposes.

\subsection{Transport}
\label{sec:transport}

The energy (per unit mass, per unit time) transported by
WIMPs can be cast as:
\begin{equation}
\label{epstrans}
    \epsilon_\mathrm{trans} = \frac{1}{4\pi r^2 \rho(r)}\frac{\mathrm{d}}{\mathrm{d}r}\big[\mathfrak{f}(K)\mathfrak{h}(r)L_\mathrm{trans, LTE}(r)\big];
    \label{eq:epstrans}
\end{equation}

with 
\begin{eqnarray}
\label{LTEtransport}
    L_\mathrm{trans, LTE}(r) = 4\pi r^2 \kappa(r)n_{\chi,\mathrm{LTE}}(r)l(r) \\
    \quad\times\,\Big[\frac{\mathrm{k}T_\star(r)}{m_\chi}\Big]^{1/2}\mathrm{k}\frac{\mathrm{d}T_\star(r)}{\mathrm{d}r} \nonumber
\end{eqnarray}
being the ``transport luminosity'' to be imputed to the 
WIMPs inside the Sun.
The Knudsen number, $K,$ indicates the ``localization'' 
of the WIMPs transport:

\begin{equation}
K=\frac{l(0)}{r_\chi} .
\label{KnudNumber}
\end{equation}

The WIMP distribution $n_{\chi,\mathrm{LTE}}$ is obtained in the 
approximation that the Knudsen number is very small and WIMPs are in
local equilibrium with the baryons:
\begin{eqnarray}
\label{LTEdens}
    n_{\chi,\mathrm{LTE}}(r) = n_{\chi,\mathrm{LTE}}(0)\Big[\frac{T_\star(r)}{T_\mathrm{c}}\Big]^{3/2} \\
    \times\exp\Big[-\int^r_0\frac{\mathrm{k}\alpha(r')\frac{\mathrm{d}T_\star(r')}{\mathrm{d}r'} + 
      m_\chi\frac{\mathrm{d}\phi(r')}{\mathrm{d}r'}}{\mathrm{k}T_\star(r')}\,\mathrm{d}r'\Big], \nonumber
\end{eqnarray}

with the normalization $n_{\chi,\mathrm{LTE}}(0,t)$ obtained imposing
$\int^{R_\star}_0 4\pi r^2 n_{\chi,\mathrm{LTE}}(r,t)\,\mathrm{d}r = N(t)$.
$T_*(r)$ is the temperature profile of the star and $\phi(r)$ is the local gravitational
potential.
The thermal diffusivity and conductivity coefficients, respectively
$\kappa (r)$ and $\alpha(r),$ are defined as:
\begin{equation}
    \alpha(r) = \sum_i \frac{\sigma_in_i(r)}{\sum_j\sigma_jn_j(r,t)}\alpha_i(\mu_i)
\end{equation}
and
\begin{equation}
    \kappa(r) = \Big\{l(r)\sum_i\big[\kappa_i(\mu_i)l_i(r)\big]^{-1}\Big\}^{-1}.
\end{equation}
with $\alpha_i(\mu_i)$ and $\kappa_i(\mu_i)$  obtained from 
tabulated values in Ref. \cite{Gould:1989hm}.
The sum runs over all the nuclear species,
$\mu_i$ is the WIMP-to-nucleus mass ratio and $n_i(r)$ is the number
density of the species $i$ at any given radius.

The two correction factors which appear in Eq. \ref{eq:epstrans}
read:
\begin{equation}
\label{radialsupp}
    \mathfrak{h}(r) \approx \Big(\frac{r-r_\chi}{r_\chi}\Big)^3 + 1
\end{equation}
and 
\begin{equation}
\label{knudsensupp}
    \mathfrak{f}(K) \approx 1 - \frac{1}{1+\mathrm{e}^{-(\ln K - \ln K_0)/\tau}} = 1 - \frac{1}{1+\big(\frac{K_0}{K}\big)^{1/\tau}}
\end{equation}
with  $\tau=0.5$ and $K_0$=0.4.

These two quantities, introduced by Gould \& Raffelt \cite{Gould:1989hm},
extend the treatment of the WIMPs energy transport obtained under
the assumption of local thermal equilibrium (i.e. small $K$),
to the case of large Knudsen numbers.
Therefore, the equations here described are virtually correct
for any value of WIMP scattering cross-sections, applying
both to local and non-local energy transport regimes.
The WIMPs energy distribution in Eq. \ref{LTEdens} differs from the analogue
expression Eq. \ref{eq:Distribuzione} valid for large $K.$
In the following, for the computation of the annihilation rate,
we adopt a distribution interpolating between the two regimes that we defined,
following \cite{Scott:2008ns}, as:
$$n_{\chi}(r)=f(K) n_{\chi,\mathrm{LTE}}(r)+\left( 1-f(K) \right) n_{\chi,nl}.$$
As commented in Sec.\ref{sec:annihilation} 
this modification is however negligible in practice.

\end{document}